\documentclass[preprint,12pt]{elsarticle}

\usepackage{amssymb}

\usepackage{color}

\journal{publication - formatting can change after accepted - }

\begin{document}

\begin{frontmatter}




\title{Crossing-effect in non-isolated and non-symmetric systems of patches}


\author[unifal]{Daniel Juliano Pamplona da Silva}
\ead{pamplona@unifal-mg.edu.br}

\address[unifal]{Universidade Federal de Alfenas - UNIFAL-MG, Rodovia Jos\'e Aur\'elio Vilela, 11.999 - 37715-400 
Po\c{c}os de Caldas, Brazil}

\begin{abstract}

The main result of this article is the determination of the minimal size for the general case of problems with two identical patches. This solution is presented in the explicit form, which allows to recuperate all the cases found in the literature as particular cases, namely, one isolated fragment, one single fragment communicating with its neighborhood, a system with two identical fragments isolated from the matrix but mutually communicating and a system of two identical fragments inserted in a homogeneous matrix. It is also addressed the new problem of a single fragment communicating with the matrix, with different life difficulty of each side. As application, it is found that the internal condition $a_{0}$ can set which system is the worst to life. This prediction confirms and extends the prediction already found in the literature between isolated and non-isolated systems.

\end{abstract}


\end{frontmatter}


\section{Introduction}
\label{Introduction}

Populations living in one \cite{Nelson1998} or more \cite{Lpez2017} fragments have been the subject of study in many natural sciences, in particular Ecology \cite{Ferraz2007}. In fact, natural scientists such as physicists \cite{Artiles2008,Kumar2011}, mathematicians \cite{Cantrell1991,Hening2017} and others \cite{Skellam1951} have been studied models that can describe, with some degree of approximation, the real phenomenon.

The study of fragmented regions problem was started by Skellam \cite{Skellam1951}, which proposed a solitary fragment which can harbor life within the patch but makes it impossible for live to exist outside it. This problem was improved by Ludwig et all \cite{Ludwig1979}. They introduced a non hard region outside the patch to the Skellam problem. Then, in the Ludwig problem, the population in study can go outside the island, but can not live there forever. The improvement of Ludwig et all brought the need of an smaller fragment than the one found by Skellam to enable stable life in the system with only one fragment. The natural sequence to Ludwig work is the introduction of another fragment in the system. Now, we have two fragments communicating by a region not favorable to life, but not infinitely hard, in such a way that is possible population elements pass from one fragment to the other. In this sense, there are previous studies for the implicit form to minimal size fragments that enables life for a more general case \cite{Pamplona2012}. The explicit form for this minimal size was presented by Kenkre and Kumar \cite{Kenkre2008} to the case where the life difficulty was the same outside the system and between the patches.

The predictions proposed above were found in one dimension, where the focus was just with the minimum length for life existence. In two dimensions, the study would concern the minimal area \cite{Azevedo2012} and it would be necessary to explore the geometry of the fragment \cite{Kenkre2008}, which would result in another scope of work, not addressed in this paper. All cases mentioned previously used \cite{Fisher1937} Fisher-Kolmogorov-Petrovskii-Piskunov (FKPP) equation to describe the behavior of a population density $u(x,t)$, in the time ($t$), moving in the space ($x$), governed by the growth rate $a$ and contained by the saturation rate $b$. In one-dimensional it is given by
%
\begin{equation} \label{FKPP}
\frac{\partial u}{\partial t} = D\frac{\partial^{2}u}{\partial x^2}+a(x)u-bu^2,
\end{equation}
%
where $D$ is the diffusion coefficient. Parameter $a$  is linked to the internal conditions of the fragment concerning life existence and parameter $b$ represents a population intraspecific competition. It defines the maximum population that can occupy a specific fragment. Parameter $b$ is closely linked to the carrying capacity.

\section{The model}
\label{The model}

\begin{figure}[ht]
\begin{center}
\includegraphics[height=3cm]{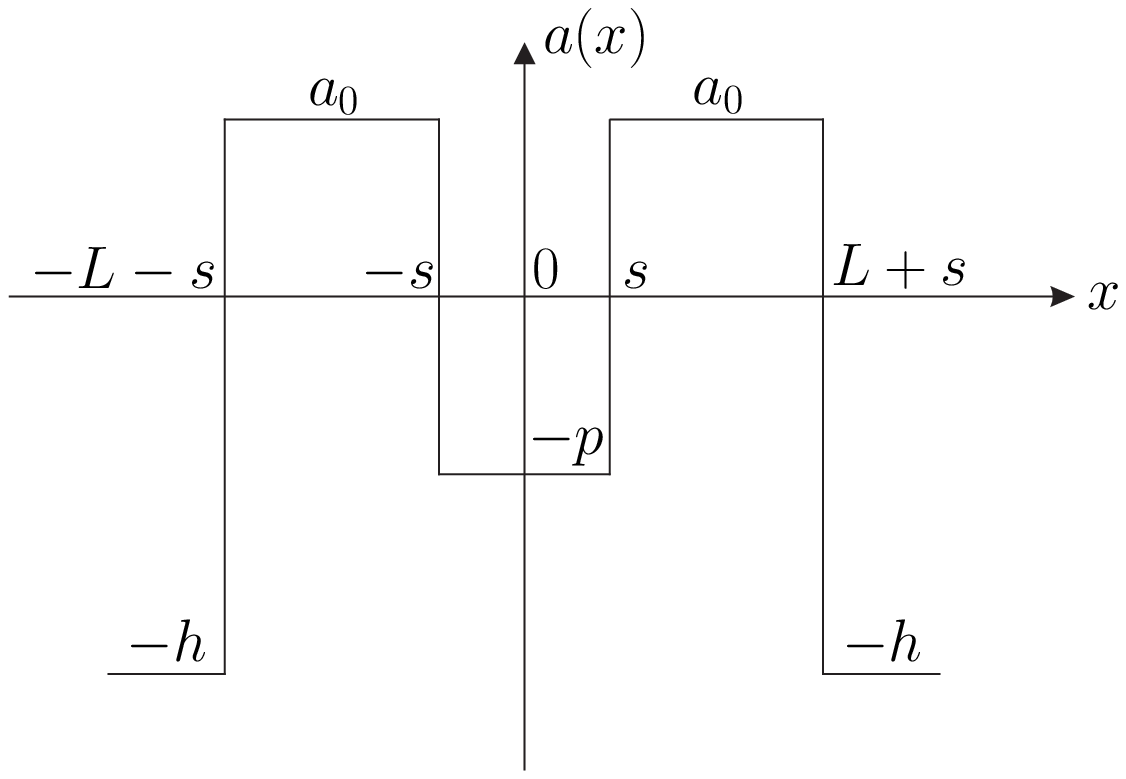}
\hspace{1cm}\includegraphics[height=3cm]{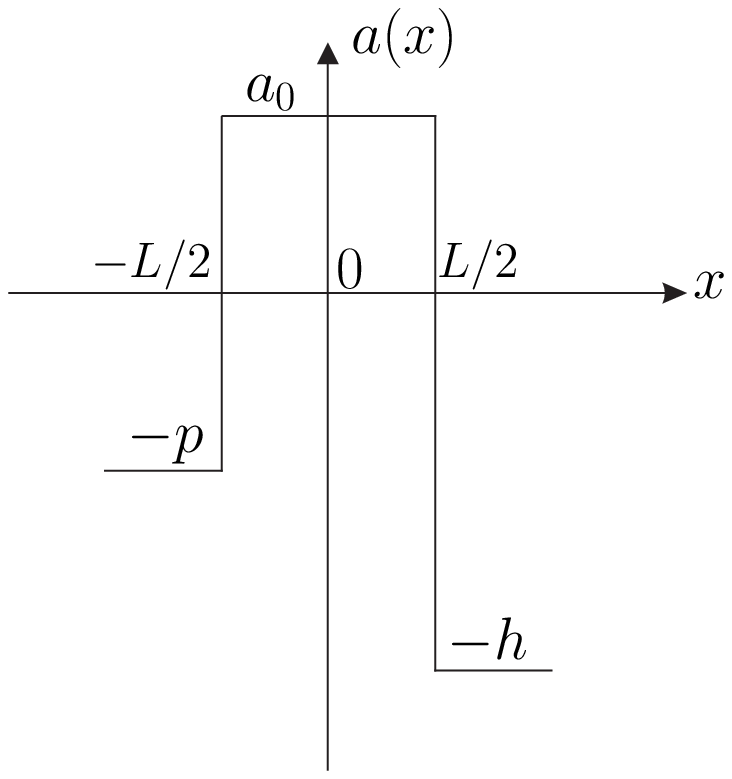}\\
\hspace{0.5cm}(a) \hspace{4cm} (b) 
\caption{\label{duplog} Representation of two identical patches with length $L$ and internal condition $a_{0}$ immersed in a matrix with life difficulty $h$, separated by a region with life difficulty $p$ and (a) of length $s$ that (b) approaches a non symmetric single patch when $s\rightarrow\infty$.}
\end{center}
\end{figure}

The article's propose is to solve the general case of two identical patches - see Fig. (\ref{duplog}a) - and recover several cases addressed in the literature. In addition, it is found the minimal size to a single non symmetric patch capable of sustaining life. The also discuss the particularities that emerge from a fragment with different condition to life difficulty of each side of the matrix - see Fig. (\ref{duplog}b).

The Eq. (\ref{FKPP}) does not have general solution for a arbitrary function $a(x)$ at any time $t$. A well known method \cite{Kenkre2008,Ludwig1979,Pamplona2012} used to find a critical condition to life existence is to suppress the nonlinear term $-u^2$. It gives a linear partial differential equation. It is then possible to use the separation of variables method, assuming that $u(x,t)=\Phi(x)\psi(t)$ and finding a ordinary differential equation (ODE) for $\psi(t)$ with respect to time and a ODE for $\Phi(x)$ with respect to space.

The resulting spatial ODE can be written as follow:

\begin{equation} \label{FKPP-le}
D\frac{d^{2}\Phi}{d x^2}+a(x)\Phi=0.
\end{equation}
%
where it has set the separation constant to zero in order to obtain the smallest fragment size that satisfy the boundary and continuity conditions.

By choosing the function $a(x)$ as a piecewise constant, the Eq. (\ref{FKPP-le}) can be solved easily in any constant piece. It describes a heterogeneous region \cite{Kenkre2008,Kenkre2003,Kraenkel2010,Pamplona2012}, where $a(x)>0$ represents a patch (life region) and $a(x)<0$ represents a dead region. Note that there are space heterogeneities, but the condition inside each patch is homogeneous, outside the conditions are homogeneous too. Then, there are only abrupt changes in the environment condition and this occurs in the frontiers between a patch and a death region. The generalization of two communicating identical patches is represented by Fig. (\ref{duplog}a).

The main mathematical propose of this paper is to find the explicit form of minimum size to each fragment, represented in Fig. (\ref{duplog}a), that enables stable life existence inside the patch. To reach the goal, it will not seek the solution to the linear equation (Eq. \ref{FKPP-le}), because it does not reflect the nonlinear equation (Eq. \ref{FKPP}). The later should be the one used as a model to describe the full problem in question. Instead, the goal is to use the linear equation to find the size of the minimal fragment, from boundary and continuity conditions, following the ideas of Ludwig et all and their successors \cite{Kenkre2008,Ludwig1979,Pamplona2017}.

These conditions are applied only for the positive values of $x$ ($x=s$ and $x=L+s$) because, from symmetry, the negative side returns the same conditions. In order to simplify the notation, it is introduced

\begin{equation}\label{galpha}
\displaystyle \alpha_{a}=\sqrt{\frac{a}{D}}, \quad \forall a.
\end{equation}

Solutions to Eq. (\ref{FKPP-le}) are:

\begin{equation}
\Phi_{I}(x)=A\cosh{(\alpha_{p}x)}, \mbox{ in region I } (-s<x<s);
\end{equation}

\begin{equation}
\Phi_{II}(x)=B\cos{(\alpha_{a_{0}} x + \phi)}, \mbox{ in region II } (s<x<L+s);
\end{equation}

\begin{equation}
\Phi_{III}(x)=Ce^{-\alpha_{h}x}, \mbox{ in region III } (x>L+s).
\end{equation}

Next, it is imposed the matching condition upon $\Phi(x)$ at each boundary. In fact:
\begin{itemize}
\item At $x=s$, the condition is given by $\Phi_{II}(s)=\Phi_{I}(s)$, then:
\begin{equation}\label{phis}
B\cos{(\alpha_{a_{0}} s + \phi)} = A\cosh{(\alpha_{p}s)}.
\end{equation}

\item At $x=L+s$, the condition $\Phi_{II}(L+s)=\Phi_{III}(L+s)$ gives:
\begin{equation}\label{philps}
B\cos{(\alpha_{a_{0}} L+\alpha_{a_{0}} s + \phi)}=Ce^{-\alpha_{h}(L+s)}.
\end{equation}

\end{itemize}

Not only the function $\Phi(x)$ should be continuous but also its derivative. Hence:

\begin{itemize}

\item In $x=s$, immediately:
\begin{equation}\label{dphis}
-B\alpha_{a_{0}}\sin{(\alpha_{a_{0}} s + \phi)} = A\alpha_{p}\sinh{(\alpha_{p}s)}.
\end{equation}
\item In $x=L+s$: 
\begin{equation}\label{dphilps}
-B\alpha_{a_{0}}\sin{(\alpha_{a_{0}} L+\alpha_{a_{0}} s + \phi)}=-C\alpha_{h}e^{-\alpha_{h}(L+s)}.
\end{equation}

\end{itemize}

Dividing Eq. (\ref{dphis}) by Eq. (\ref{phis}), it follows:
\[
-\alpha_{a_{0}}\tan{(\alpha_{a_{0}} s + \phi)} = \alpha_{p}\tanh{(\alpha_{p}s)} \Rightarrow
\]
\begin{equation}\label{step1}
-\alpha_{a_{0}} s - \phi=\arctan{\left[\frac{\alpha_{p}}{\alpha_{a_{0}}}\tanh{\alpha_{p}s}\right]}.
\end{equation}

Dividing Eq. (\ref{dphilps}) by Eq. (\ref{philps}), it results:
\[
-\alpha_{a_{0}}\tan{(\alpha_{a_{0}}L + \alpha_{a_{0}}s + \phi)} = -\alpha_{h} \Rightarrow
\]
\begin{equation}\label{step2}
\alpha_{a_{0}} L+\alpha_{a_{0}} s + \phi=\arctan{\left(\frac{\alpha_{h}}{\alpha_{a_{0}}}\right)}.
\end{equation}

Eqs. (\ref{step1}) and (\ref{step2}) can be added. The convenient form of the result from this operation is:

\begin{equation}\label{step3}
L=\frac{1}{\alpha_{a_{0}}}\left\{\arctan{\left(\frac{\alpha_{h}}{\alpha_{a_{0}}}\right)}+\arctan{\left[\frac{\alpha_{p}}{\alpha_{a_{0}}}\tanh{\alpha_{p}s}\right]}\right\},
\end{equation}
which can be rewritten in terms of the initial variables as

\begin{equation}\label{eqLph}
L_{ph}=\sqrt{\frac{D}{a_{0}}}\left\{\arctan{\left(\sqrt{\frac{h}{a_{0}}}\right)}+\arctan{\left[\sqrt{\frac{p}{a_{0}}}\tanh{\left(s\sqrt{\frac{p}{D}}\right)}\right]}\right\}.
\end{equation}

Subscript $ph$ is only to identify the general expression for minimal size of one patch in a system of two identical patches - Eq. (\ref{eqLph}) - as well as the others subscripts that appear below to identify the different particular cases of minimal size $L$.

\section{Results and Analysis}
\label{Results and Analysis}

The expression (\ref{eqLph}) is one of the main results in this work. It is a generalization of all the problems of systems with one patch or two identical patches. It is possible to recover all the particular cases of literature from this equation. If $h$ and $p$ go to infinity, Skellam \cite{Skellam1951} result is recovered. If a finite value is chosen to $h=p$ and $s\rightarrow\infty$, Ludwid et all formula \cite{Ludwig1979} is recovered. Choosing $p$ and $s$ as any finite values and $h\rightarrow\infty$, Pamplona et all \cite{Pamplona2017} prediction is obtained. Choosing finite values to $s$ and $h=p$, Kenkre and Kumar \cite{Kenkre2008} result is recovered. See the resume in table \ref{results}.

\begin{table}[ht]
\begin{center}
{\footnotesize
\caption{\label{results} Particular cases for minimum size of patches found in the literature.}
\begin{tabular}{|c|c|c|}
\hline  
Problem & Minimum size of fragments & Choice\\ 
\hline
$\begin{array}{c} \mbox{Skellam}\\ \mbox{\cite{Skellam1951}} \end{array}$ & $\displaystyle L_{si}=\pi{\sqrt{\frac{D}{a_{0}}}}$ & $\begin{array}{l} { }\\ p\rightarrow\infty\\h\rightarrow\infty \end{array} $\\
\hline 
$\begin{array}{c} \mbox{Ludwig}\\ \mbox{\cite{Ludwig1979}} \end{array}$ & $\displaystyle L_{sn}=2\sqrt{\frac{D}{a_{0}}} \arctan{\sqrt{\frac{h}{a_{0}}}}$  & $\begin{array}{l} { }\\h=p\\s\rightarrow\infty \end{array} $\\
\hline 
$\begin{array}{c} \mbox{Pamplona}\\ \mbox{\cite{Pamplona2017}} \end{array}$ & $\displaystyle L_{di}=\sqrt{\frac{D}{a_{0}}}\left\{\frac{\pi}{2}+\arctan{\left[\sqrt{\frac{p}{a_{0}}}\tanh{\left(s\sqrt{\frac{p}{D}}\right)}\right]}\right\}$ & $\begin{array}{l} { }\\ h\rightarrow\infty \\ { } \end{array} $\\
\hline 
$\begin{array}{c} \mbox{Kenkre}\\ \mbox{\cite{Kenkre2008}} \end{array}$& $ \displaystyle L_{dh}=\sqrt{\frac{D}{a_{0}}}\left\{\arctan{\sqrt{\frac{h}{a_{0}}}}+\arctan{\left[\sqrt{\frac{h}{a_{0}}}\tanh{\left(s\sqrt{\frac{h}{D}}\right)}\right]}\right\}$ & $\begin{array}{l} { } \\ h=p \\ { } \end{array} $\\
\hline 
\end{tabular}\\}
\end{center}
\end{table}

After re-obtaining several known cases on the literature with the general result, let us turn to the new problem: a single patch with different boundary conditions on each end, i.e. a problem without a symmetric form, such as the one presented in Fig. (\ref{duplog}b). For this case, the minimal size can be obtained directly from Eq. (\ref{eqLph}) by taking the limit $s\rightarrow\infty$, namely,
\begin{equation}\label{eqLsph}
L_{sph}=\sqrt{\frac{D}{a_{0}}}\left\{\arctan{\left(\sqrt{\frac{h}{a_{0}}}\right)}+\arctan{\left(\sqrt{\frac{p}{a_{0}}}\right)}\right\}.
\end{equation}

With the Eq. (\ref{eqLsph}) it is possible to obtain the minimal size for a semi isolated fragment taking the limit $h\rightarrow\infty$. This result is:
\begin{equation}\label{eqLspi}
L_{spi}=\sqrt{\frac{D}{a_{0}}}\left\{\frac{\pi}{2}+\arctan{\left(\sqrt{\frac{p}{a_{0}}}\right)}\right\}.
\end{equation}

\begin{figure}[ht]
\begin{center}
\includegraphics[height=9cm]{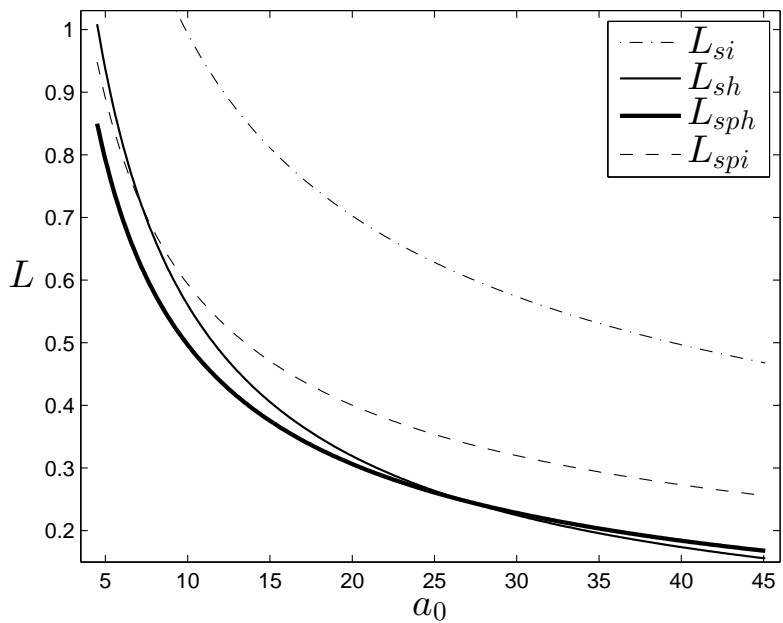}
\caption{\label{crosseffect}  Plots of minimum size of only one patch versus growth rate ($a _{0}$) for the following cases: the both sides of the patch are isolated ($L_{si}$); one side is isolated and the other one communicates with the matrix ($L_{spi}$); the both sides are non isolated and have the same external life difficulty ($L_{sn}$, $h=15$) and the both sides are non isolated and have different external life difficulty ($L_{sph}$, $h=100$) at each side. The plots to $L_{spi}$ and $L_{sph}$ use the same value $p=1$.}
\end{center}
\end{figure}

Eq. (\ref{eqLspi}) is in accordance with the previous prediction in \cite{Pamplona2017} as well as with Ludwig et all formula (Table \ref{results}). This result also enables the confirmation of Pamplona et all \cite{Pamplona2017} prediction about the isolation effects. Moreover, it extends the validity of this phenomenon to non isolated systems - see Fig. (\ref{crosseffect}).

Fig. (\ref{crosseffect}) shows the crossing of $L_{sh}$ curve with $L_{spi}$ and $L_{sph}$ curves. The crossing of $L_{spi}$ with $L_{sh}$ confirms the result of Pamplona et all \cite{Pamplona2017}: the semi-isolation leads to a high dependence of the minimum fragment size on its internal conditions. This result was found by Pamplona et all for a system of two patches with a relief in the region between them. The novelty presented here is the elimination of the parameter $s$, the length of separation between patches. Now there is no doubt that this effect does not need the separation between the patches. Actually, it does not need the existence of two (neighbors) patches.

For the crossing-effect to appear, it is necessary a relief on some end, but the other boundary does not need to be isolated. The crossing of $L_{sh}$ and $L_{sph}$ in Fig. (\ref{crosseffect}) shows this clearly. In other words, for this effect to appear, it is required fragment $L_{sph}$ to have one side with a relief ($p=1$ - small life difficulty) and the other side with a big life difficulty ($h=100$) while the other fragment has to have the both sides with middle life difficulty ($h=15$). In the last case, both sides don't need to have the same life difficulty condition. For example, crossing-effect exists between two $L_{sph}$ curves, one with $h = 100$ and $p = 1$ and other one with $h = 20$ and $p = 10$. The curve $L_{sph}$ with $h = 20$ and $p = 10$ was not plotted in Fig. (\ref{crosseffect}) because it would be very close to the $L_{sn}$ curve with $h = 15$.

The case of only one isolated fragment is yet the worst to life existence, once it needs the larger size of fragment in all cases. This is because it is not possible to insert a relief in any part without the loss of the system properties. In the case of an isolated system with two patches, the relief is the region between them.

\section*{Conclusion}
\label{Conclusion}

This paper confirms the break down \cite{Pamplona2017} of the paradigm \cite{Kenkre2008,Ludwig1979,Pamplona2012} that isolated systems is always worst to life than a non isolated systems. It also introduces a new conception for the analysis of minimal size of fragments, namely, a fragment can be bigger or smaller than other one for a specific set of external conditions of fragments, depending only on the internal condition of the fragments, that is the same for both. In other words: take a patch 1 (or system of patches) with a set 1 of parameters to external conditions and internal condition $a_{0}$ that need size $L_{1}$ to enable stable life inside it. Take a patch 2 (or system of patches) with a different set 2 of parameters and the same internal condition of patch 1 ($a_{0}$), that need size $L_{2}$ to enable stable life inside it. If specific sets of parameters (set 1 and set 2) are chosen. Then, $L_{1}$ is bigger than $L_{2}$ for small values of $a_{0}$ and $L_{1}$ is smaller than $L_{2}$ for high values of $a_{0}$.

This result was found in a previous paper \cite{Pamplona2017} for isolated system of two fragments with a relief in the difficulty life condition - the region between the patches. Here, this result is expanded for fragments semi-isolated or totally non-isolated of the external region, leading to the conclusion that this phenomenon is not an effect of isolation, although the effect is linked to the disparity of the external conditions. This extension is possible due to Eq. (\ref{eqLph}), the central analytic result of this paper, which generalizes a system of two identical fragments and presents the minimal size for one non-symmetric fragment as a particular case - Eq. (\ref{eqLsph}).

The existence of the crossing-effect in systems of finite external conditions introduces a bigger applicability and facility of verification in the phenomenon, once it is not simple to find or construct real systems with infinite parameters, because infinite is, almost always, a mathematical concept with difficult implementation.

The high dependence of the minimum fragment size on its internal condition brings up concern on this condition in the context of extinction, stable life in a patch or other effects linked to criticality context.

\section*{Acknowledgments}
\label{Acknowledgments}
The author is grateful to PET - Programa de Educa\c{c}\~ao Tutorial for financial support, Rodrigo Rocha Cuzinatto for text revision and Lorielen Calixto Ramos and Renato Pacheco Villar for fruitful discussions.



\end{document}